\documentclass[pra,twocolumn,showpacs,preprintnumbers,amsmath,amssymb]{revtex4}
\usepackage[dvips]{graphicx}% Include figure files
\usepackage{dcolumn}% Align table columns on decimal point
\usepackage{bm}% bold math
\usepackage{braket}

\begin{document}

\preprint{}

\title{Faraday Rotation with Single Nuclear Spin Qubit in a High-Finesse Optical Cavity}

\author{Nobuyuki Takei$^{1,\dag}$, Makoto Takeuchi$^{1,\ddag}$, Yujiro Eto$^{1,2}$\\
Atsushi Noguchi$^{1,2}$, Peng Zhang$^{1}$, Masahito Ueda$^{1,3}$, and Mikio Kozuma$^{1,2}$}

\affiliation{%
$^{1}$ERATO Macroscopic Quantum Control Project, JST, 2-11-16 Yayoi, Bunkyo-Ku, Tokyo 113-8656, Japan}

\affiliation{%
$^{2}$Department of Physics, Tokyo Institute of Technology, 2-12-1 O-okayama, Meguro-ku, Tokyo 152-8550, Japan}

\affiliation{%
$^{3}$Department of Physics, University of Tokyo, Hongo, Bunkyo-ku, Tokyo 113-0033, Japan}

\altaffiliation[Present affiliation:]{$\dag$ Institute for Molecular Science,
$\ddag$ National Institute of Information and Communications Technology}

\date{\today}
%\keywords{Quantum computation architectures and implementations, 
%Cavity quantum electrodynamics}

\begin{abstract}
When an off-resonant light field is coupled with atomic spins, its polarization can rotate depending on the direction of the spins via a Faraday rotation which has been used for monitoring and controlling the atomic spins.
We observed Faraday rotation by an angle of more than $10$~degrees for a single 1/2 nuclear spin of ${}^{171}\mathrm{Yb}$ atom in a high-finesse optical cavity. By employing the coupling between the single nuclear spin and a photon, we have also demonstrated that the spin can be projected or weakly measured through the projection of the transmitted single ancillary photon.
\end{abstract}

\maketitle

\section{Introduction}

Faraday rotation is a phenomenon in which the polarization of a light field rotates depending on the spin direction in an optical medium, e.g., an atomic system. Because this interaction deterministically entangles atoms with photons, it has been extensively investigated in the field of quantum information processing~(QIP)~\cite{Julsgaard01}. The Faraday rotation has also been utilized to perform quantum nondemolition measurements of the collective spins of an atomic ensemble~\cite{Kuzmich98,Takahashi99}, followed by the demonstration of spin squeezing~\cite{Julsgaard01,Takano09,Kitagawa93}. This rotation arises from the phase shift acquired by photons via dispersive interactions. Even with single atoms, large phase shifts have been observed in a high-finesse optical cavity~\cite{Turchette95} and in a dipole trap with a tightly focused probe beam~\cite{Aljunid09}. Such a conditional phase shift and polarization rotation based on the atomic state provide a key building block in QIP~\cite{NC}, leading to atom-photon entanglement or a mediator for photon-photon entanglement.

Through the entanglement formation, the photon field serves as an ancilla for monitoring and controlling the primary atomic system. The Faraday rotation interaction between the photon field and atomic system and the subsequent projective measurement on the ancilla photons can constitute an ancilla-assisted measurement~\cite{Hume07} on the atoms. The resultant change in the atomic state after the measurement is described by the Kraus measurement operators~\cite{NC}. This measurement framework provides a rich variety of controllability in quantum measurements, including feedback control of quantum state reduction~\cite{Handel05}, reversible measurement~\cite{Ueda92,Royer94,Terashima06,Katz08}, and error correction~\cite{Koashi99}.

In this article, we report the observation of Faraday rotation by an angle of more than $10$~degrees for a single 1/2 nuclear spin of the ${}^{171}\mathrm{Yb}$ atom. The nuclear spin is an ideal candidate for a quantum bit~(qubit) because of its long coherence time \cite{Kane98,Childress06}. In our experiment, the spin-photon coupling is greatly enhanced by using a high-finesse optical microcavity. In the present work, we have also demonstrated ancilla-assisted quantum measurement. The spin state is projected or weakly measured through polarization-dependent single photon counting for a weak coherent state of a probe pulse, which can be used for implementing the error correction process~\cite{Koashi99}.

Note that our cavity quantum electrodynamics~(QED) system is also applicable for constructing a deterministically controlled NOT gate in which control and target qubits are represented by a nuclear spin state and a polarization state of a photon, respectively. Depending on the spin direction, the polarization rotates to different directions; +~(-) $45$~degrees for the down~(up) spin in an ideal case. By rotating the polarization by $45$~degrees after the cavity, the output polarization remains unchanged compared to the original one when the nuclear spin is up. On the contrary, when the spin is down, the polarization rotates by $90$~degrees.

This paper is organized as follows. In Sec.~II, we briefly describe our experimental apparatus and the method for real-time selection of a spin state strongly interacting with the cavity field. In Sec.~III, we show the experimental results for the Faraday rotation. In Sec.~IV we present ancilla-assited measurement in which the spin direction changes based on the measurement result on the ancilla photon. In Sec.~V, we discuss how to achieve a larger Faraday rotation angle. In Sec.~VI, we conclude this paper.

\section{Experimental procedure}

\subsection{Apparatus}

\begin{figure}[htbp]
\begin{center}
\scalebox{0.16}{\includegraphics{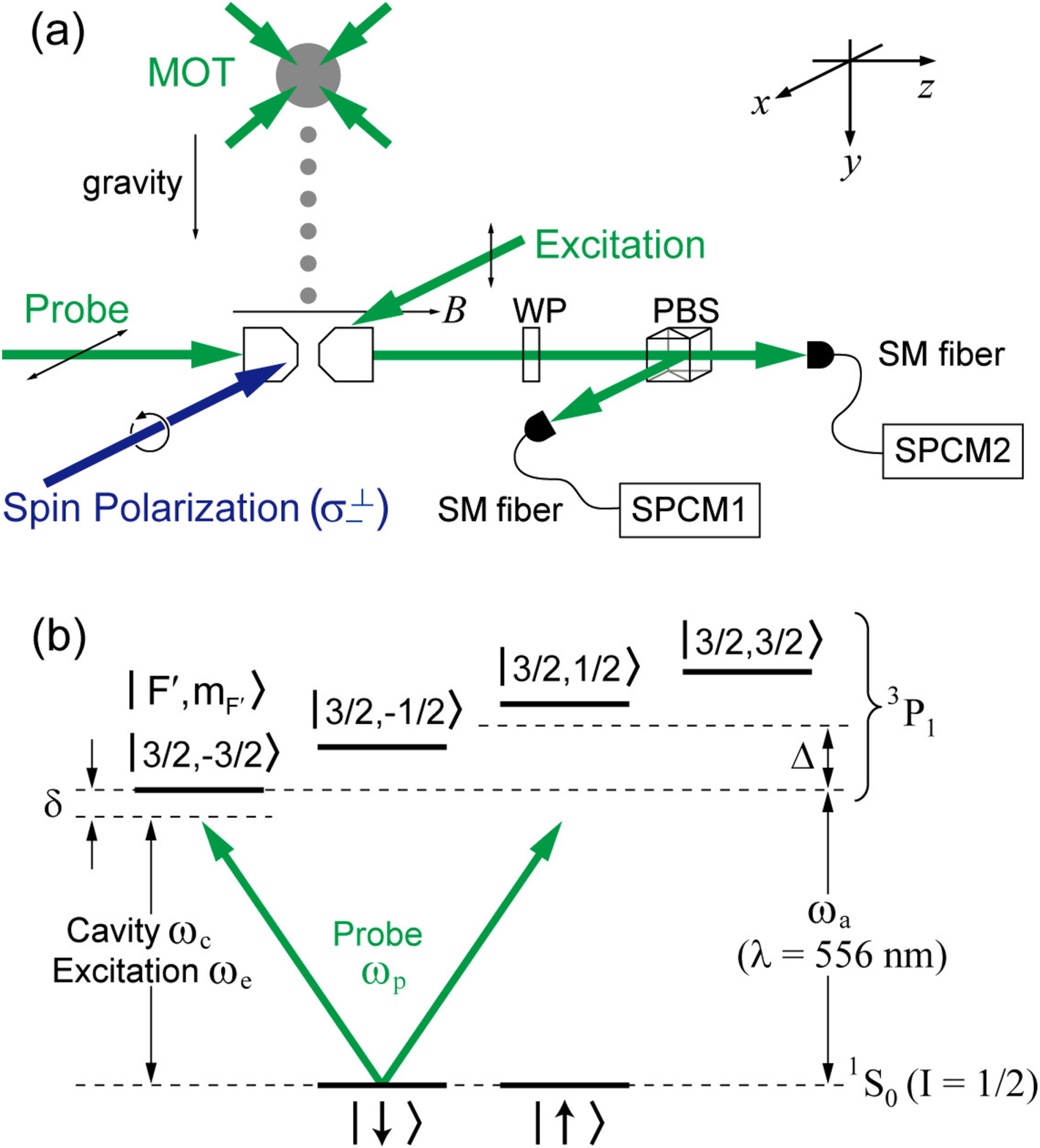}}
\caption{(color online)
(a) Experimental configuration. WP, wave plate; PBS, polarizing beam splitter; SM fiber, single mode fiber; SPCM, single photon counting module.
(b) Energy-level diagram of $\mathrm{^{171}Yb}$. The magnetic substates $m_I=+1/2$ and $-1/2$ in the ground state $^1\mathrm{S}_0(I=1/2)$ are denoted by $\ket{\uparrow}$ and $\ket{\downarrow}$, respectively. The substates in the excited state $^3\mathrm{P}_1$ are labeled as $\ket{F',m_{F'}}$. The Zeeman shift of the $\ket{3/2, -3/2}$ state due to the bias magnetic field is $\Delta=2\pi\times 71~$MHz.
}
\label{energy}
\end{center}
\end{figure}
Our experimental setup and the relevant energy levels are shown in Fig. 1. 
We briefly describe our apparatus here (details are described in Ref.~\cite{Takeuchi09}).
First, $\mathrm{^{171}Yb}$ atoms are prepared in a magneto-optical trap~(MOT) with $^1\mathrm{S}_0$$-$$^3\mathrm{P}_1$ intercombination transition~(wavelength $\lambda=556$~nm), which is situated 7~mm above the Fabry-Perot microcavity.
The atoms are then released and introduced into the cavity under gravity.
During freefall for 40~ms, a bias magnetic field of 34~G is switched on along the cavity axis to set the quantization axis.
The cavity consists of two identical concave mirrors with a radius of curvature 50~mm.
These mirrors are separated by a gap of $150~\mu$m and the waist size is $w_0=19~\mu$m.
The MOT loading time is adjusted so that the intra-cavity atom number becomes much less than unity.
The average transit time in the cavity is approximately 120~$\mu$s.
The resonant frequency of the cavity $\omega_c$ is stabilized to be near resonant to the transition $\ket{\downarrow}\leftrightarrow \ket{3/2, -3/2}$ with frequency $\omega_a$. 
Here $\ket{\uparrow}$ and $\ket{\downarrow}$ denote the magnetic substates $m_I=+1/2$ and $-1/2$ in the ground state $^1\mathrm{S}_0(I=1/2)$, respectively. The magnetic sublevels in the excited state $^3\mathrm{P}_1$ are labeled as $\ket{F',m_{F'}}$. The Zeeman shift of the $\ket{3/2, -3/2}$ state due to the bias magnetic field is $\Delta=2\pi\times 71~$MHz.
Our system is characterized by the following three parameters: the maximum interaction rate between atoms and photons $g_0=2\pi\times2.8$~MHz, cavity decay rate (HWHM of the cavity resonance line) $\kappa=2\pi\times4.5$~MHz, and atom decay rate (natural linewidth) $\gamma=2\pi\times 182~$kHz.
The input beam is coupled to the cavity with an efficiency of 0.6, and the escape efficiency is 0.9.
The output from the cavity is split using a polarizing beam splitter~(PBS), and each output from the PBS is coupled to a single-mode fiber~(SM fiber) with an efficiency of 0.7,  and it is detected using a single photon counting module~(SPCM) 1, 2~ (PerkinElmer SPCM-AQR-14-FC), whose detection efficiency is 0.6 at 556~nm. The total detection efficiency of a photon emitted from an atom is $\eta=0.4$.
The wave plate~(WP) before the PBS is selected from a half~(HWP) or quarter one~(QWP) depending on the purpose.
When the released atomic cloud reaches the cavity, the locking beam that stabilizes the cavity length is turned off for $3$~ms to avoid the SPCMs being saturated by such an intense beam.

\subsection{Real-time state selection}

By shining an excitation beam from the x-axis (see Fig.1(a)) and detecting fluorescence photons in the output mode, an atom passing through the center of the cavity mode is selected in real time in the following manner~\cite{Takeuchi09}.
The excitation beam is nearly resonant to the atomic transition with frequency $\omega_e~(\simeq\omega_a)$. The beam waist is 24~$\mu$m and the typical power is 300~nW at $\omega_e=\omega_a$. The polarization of the beam is linear along the y-axis and it can be decomposed into $\sigma_+$ and $\sigma_-$ components for the quantization axis~(z-axis). The $\sigma_-$ component with the Rabi frequency $\Omega$, typically $2\pi\times1.4$~MHz, excites only a $\ket{\downarrow}$ atom to the $\ket{3/2, -3/2}$ state. 
The atom decays back to the $\ket{\downarrow}$ state in a cyclic manner, emitting a photon into the cavity mode.
Around the center of the mode, where an atom interacts with the field more strongly, the emission cycle becomes shorter due to the Purcell effect~\cite{Kleppner81}.
The maximum emission rate to the output is given by $\kappa \Omega^2/4g_0^2$ under the condition of $g_0^2\gg\kappa\gamma$, which was calculated as $1.9\times10^{6}~\mathrm{s}^{-1}$.
By taking into account the detection efficiency $\eta=0.4$, the maximum photon counting rate was estimated to be $7.6\times10^5~\mathrm{s}^{-1}=1/(1.3~\mu\mathrm{s})$ typically.
Based on the electrically added signals from two SPCMs, if two photons are detected within $600$~ns, which we call coincidence, the atom emitting the photons is considered to have passed through the center of the cavity mode. We made the coincidence window rather shorter than $1.3~\mu\mathrm{s}$ so that only the atom around the center was surely selected.
Note that the excitation by the $\sigma_+$ component is negligible because of the large detuning.
Moreover, when an atom is in the $\ket{\uparrow}$ state, its excitation is also negligible for the same reason.
Therefore, a single atom in the $\ket{\downarrow}$ state can be selected through coincidence detection.
\begin{figure}[tbp]
\begin{center}
\scalebox{0.13}{\includegraphics{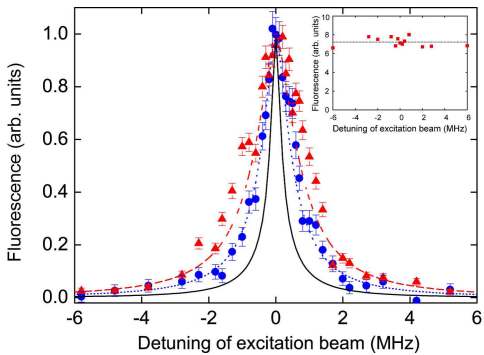}}
\caption{(color online)
Normalized fluorescence as a function of the detuning of the excitation beam. Blue data points~(circle): the power of the beam is 100~nW. Red points~(triangle): 300~nW. Solid black curve: 1~nW. Dotted blue curve: 100~nW. Dashed red curve: 300~nW. All curves are calculated based on a modified model~\cite{Turchette95a}. The inset shows the fluorescence taken with the power adjusted at each detunig.}
\label{flu}
\end{center}
\end{figure}

When the detuning of the excitation beam is larger, the excitation probability decreases, and therefore the flux of fluorescence photons for coincidence decreases. 
To keep the coincidence rate constant even at larger detuning, we increased the power of the excitation beam and also used power broadening.
Figure 2 shows the fluorescence from single atoms as a function of the detuning of the excitation beam. The detected photon counts were accumulated without the coincidence method. The theoretical curves were obtained based on a model modified from that described in Ref.~\cite{Turchette95a}.
Here, we assumed that the atoms dropped randomly onto the intersection of the excitation beam and the cavity field, and we took an average of all the calculated results.
The solid curve with a power of 1~nW indicates that the broadening is purely due to the Purcell effect.
Other curves involve power broadening, and these are in good agreement with our results. The inset of Fig.~2 shows that the adjusted fluorescence is almost constant over an entire range of detuning. The power of the excitation beam ranged from 300 nW at resonance to 18 $\mu$W at the detuning of $\pm$6~MHz.

\section{experiments}

\subsection{Faraday rotation}

After the real-time selection by coincidence, we perform the subsequent procedure of Faraday rotation within a time window of approximately $30~\mu$s~(Fig. 3~(a)).
During this window, the atom is close to the mode axis.
A linearly polarized probe pulse is sent to the cavity, where the photon number in the empty cavity is typically tuned to be 0.01. The frequency $\omega_p$ of the probe is set to be the same as the excitation beam frequency and the cavity resonant frequency, i.e. $\omega_p=\omega_e=\omega_c$, and it is also detuned from the atomic resonance as $\delta=\omega_p-\omega_a$.
\begin{figure}[tb]
\begin{center}
\scalebox{0.17}{\includegraphics{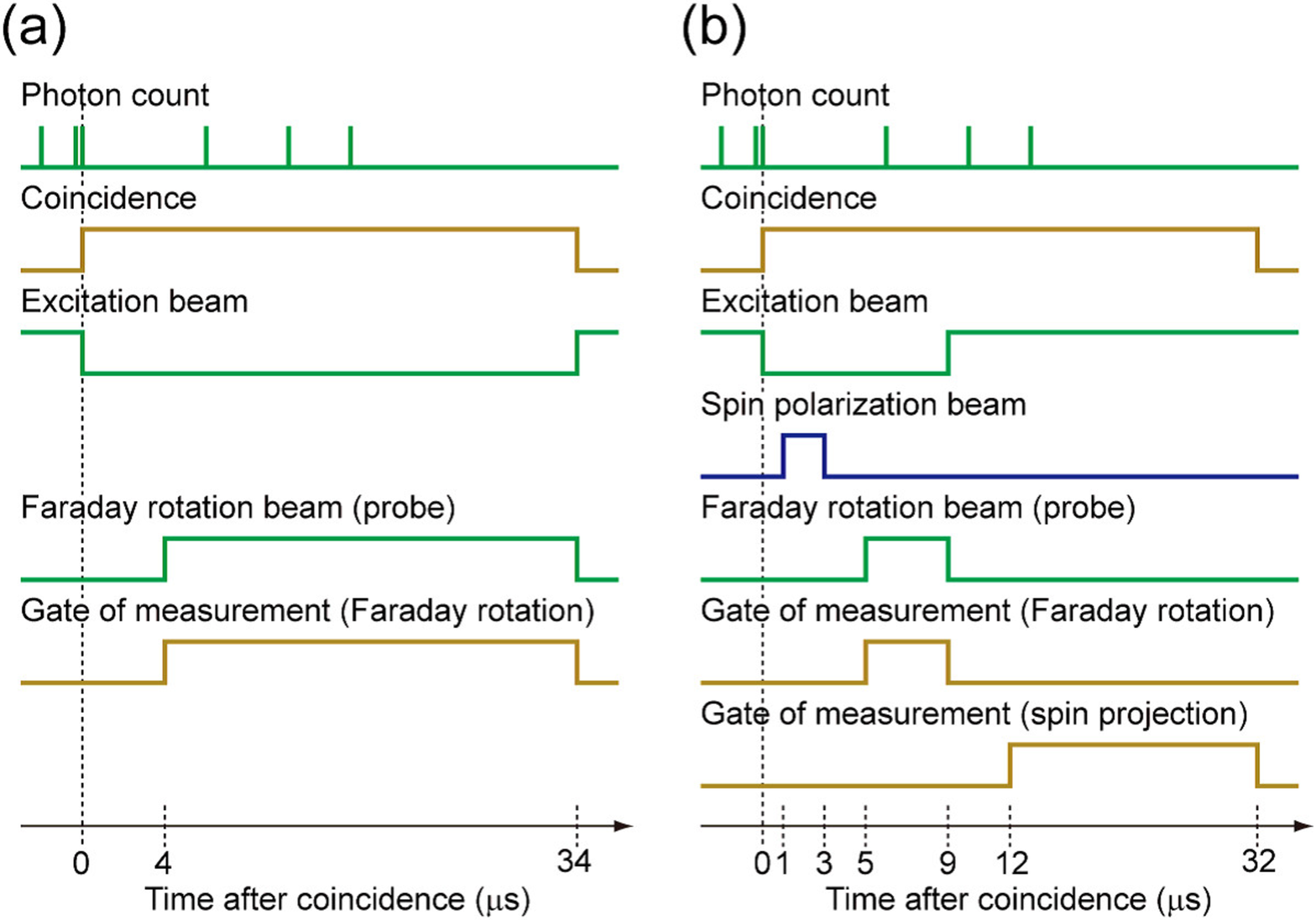}}
\caption{(color online)
Time chart of the experimental procedure for the measurements of (a) Faraday rotation and (b) variation of spin polarization.
}
\label{chart}
\end{center}
\end{figure}
We observe the dispersive interaction between the $\sigma_-$ component of the pulse and an atom in the selected $\ket{\downarrow}$ state.
As in the case of the excitation, the interactions between other combinations are inhibited.
Considering the atom-photon interaction under the weak driving condition, the transmittance of the $\sigma_-$ component for the $\ket{\downarrow}$ state is given by~\cite{Turchette95a}
\begin{equation}
T_-(\delta)=\frac{\kappa(\gamma/2-i\delta)}{\kappa(\gamma/2-i\delta)+g(r)^2},
\end{equation}
and $T_+(\delta)=1$ for the $\sigma_+$ component.
Here, the transmittance is normalized by that without atoms.
The interaction term~$g(r)$ is a function of the location of the atom: $g(r)=g_0 \exp \left[ -(x^2+y^2)/{w_0^2} \right] \cos (2\pi z/\lambda)$.
In general, $T_-(\delta)$ is complex, and therefore the polarization of the output photons becomes elliptic.

Figure 4~(a) shows the absolute value of the transmittance for the $\sigma_-$ component $|T_-(\delta)|$.
A QWP is placed behind the cavity, so that the $\sigma_-$ component is reflected from the PBS and detected using SPCM1. 
The measured data is normalized to the $\sigma_+$ beam transmitted into SPCM2~($T_+(\delta)=1$), and again normalized by the transmittance without atoms.
The theoretical solid curve is obtained by the following model.
While the coincidence method selects atoms exhibiting almost the maximum value of $g$ , the atoms travel due to thermal motion at $40~\mu$ K in the MOT and free fall~(y-axis), where a root-mean-squared velocity is 0.04~m/s and the free fall velocity is $v_y=0.3$~m/s.
Assuming $\sqrt{v_x^2+v_z^2}=0.04$~m/s during the measurement time of 34~$\mu$s, we take the average of all the possible trajectories.
The curve is in good agreement with the measured results without fitting.
\begin{figure}[tbp]
\begin{center}
\scalebox{0.23}{\includegraphics{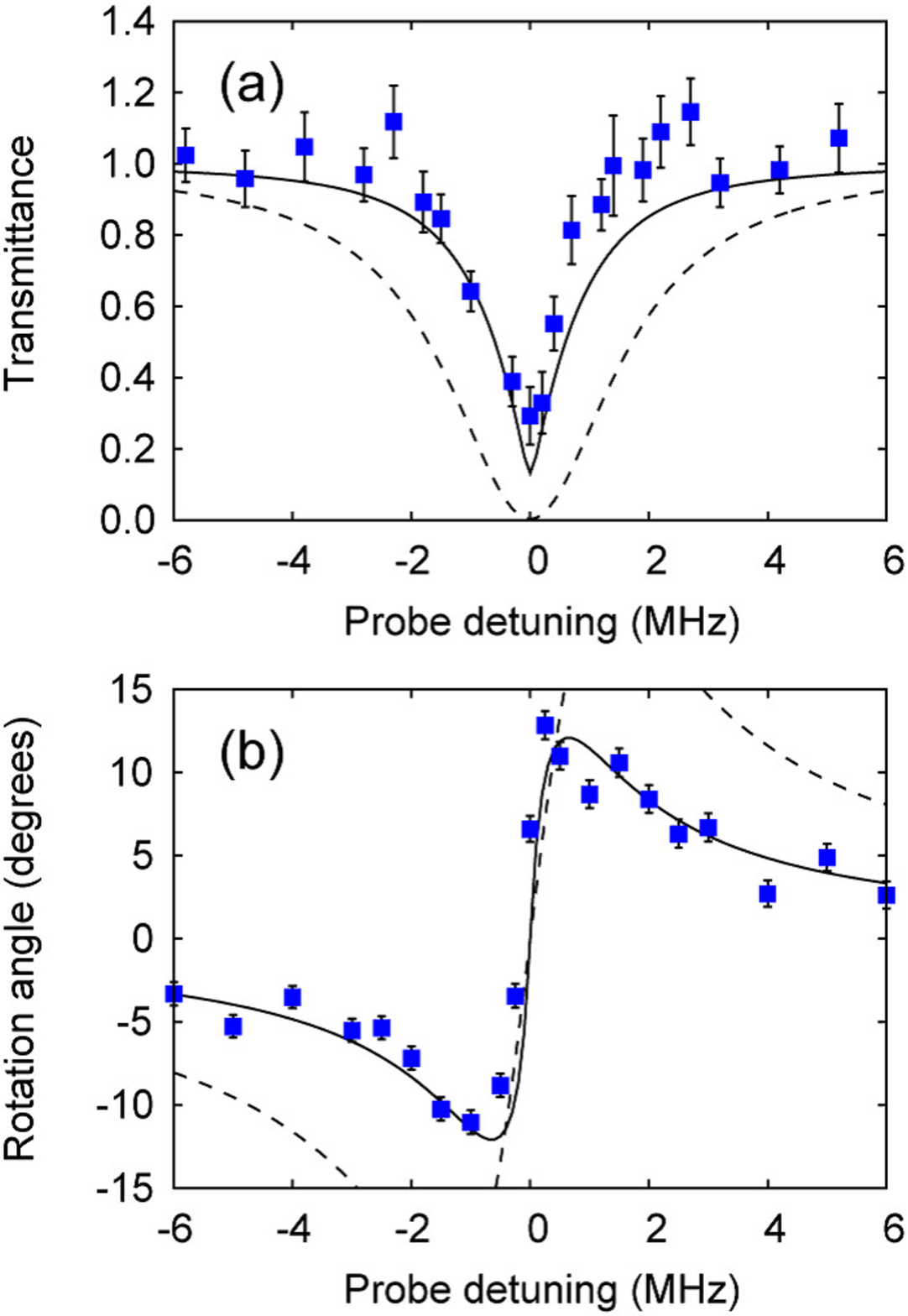}}
\caption{(color online)
(a) Transmittance for the $\sigma_-$ component of the probe pulse and (b) rotation angle of polarization for the transmitted probe as a function of the detuning. Solid curves are calculated without fitting based on our model (see text). Dashed curves are obtained with the maximum coupling $g_0$.
}
\label{property}
\end{center}
\end{figure}

The measured results for Faraday rotation, i.e., the polarization rotation of the transmitted probe, are shown in Fig.~4~(b).
The QWP was replaced with a HWP and it was adjusted so that two output powers from the PBS were balanced in the absence of atoms.
The ratio between the photon counts at the two SPCMs provides us with information about polarization angle,
\begin{equation}
\varphi=\arccos\left(\sqrt{\frac{n_{T}}{n_{T}+n_{R}}}\right)-\pi/4.
\end{equation}
Here, $n_{T}$ and $n_{R}$ are the numbers of transmitted and reflected photons, respectively, at the PBS.
We measured the value of $\varphi$ alternately with and without atoms, $\varphi_{\mathrm{w}}$ and $\varphi_{\mathrm{wo}}$, and then normalized these values to obtain the resultant rotation angle $\theta=\varphi_{\mathrm{w}}-\varphi_{\mathrm{wo}}$.
The theoretical curves were derived using a master-equation approach~\cite{Carmichael}.
Again the atomic motion was taken into account as mentioned above.
A Faraday rotation by an angle of more than $10$~degrees was observed at the detuning of approximately $\pm$1~MHz.
Due to the atomic motion, the rotation angle observed became almost half of that calculated with the maximum coupling $g_0$. If the trapping of the atom inside the cavity is achieved~\cite{McKeever03,Nubmann05}, such motion will be suppressed so that the rotation angle is expected to increase.

\subsection{Ancilla-assisted measurement: Variation of spin polarization}

Through the Faraday rotation, the primary quantum system (nuclear spin) is coupled to the ancillary system (photon), and therefore, one can perform the ancilla-assisted measurement. 
To analyze the results of our measurement, we start with a simplified model of pure rotation. Suppose that the spin is prepared in a superposition state $\ket{\text{spin}}_\mathrm{s}=\alpha \ket{\uparrow}_\mathrm{s} + \beta \ket{\downarrow}_\mathrm{s}$ and the polarization of the incident photon is parallel to the x-direction.
We represent the state of polarization by angle $\xi$ from the x-axis as $\ket{\xi}_\mathrm{p}$, so that the incident photon is in the $\ket{0}_\mathrm{p}$ state. After the Faraday rotation, the total state becomes $\alpha \ket{\uparrow}_\mathrm{s}\ket{0}_\mathrm{p} + \beta \ket{\downarrow}_\mathrm{s}\ket{\theta}_\mathrm{p}$, where $\theta$ is the rotation angle obtained for the $\ket{\downarrow}_\mathrm{s}$ state and its amplitude and sign can be controlled by changing the probe detuning $\delta$. 
Because the HWP is placed in front of the PBS, the detection of the photon at the transmission side of the PBS corresponds to the projection with the specific polarization basis $\ket{\phi}_\mathrm{p}$, and the resultant spin state is given by $\alpha\,\,_\mathrm{p}\!\!\braket{\phi|0}\!{}_\mathrm{p}\ket{\uparrow}_\mathrm{s}+ \beta\,\,_\mathrm{p}\!\!\braket{\phi|\theta}\!{}_\mathrm{p}\ket{\downarrow}_\mathrm{s}$. In the case of the detection at the reflection side, $\phi$ is replaced with $\phi+\pi/2$.
Depending on the measurement basis and the result of the measurement, the final spin state changes, which is represented by a set of measurement operators $\{\hat{M}(\theta)_{\phi},\hat{M}(\theta)_{\phi+\pi/2}\}$, where $\hat{M}(\theta)_{\phi}={}_\mathrm{p}\!\!\braket{\phi|0}\!{}_\mathrm{p}\ket{\uparrow}_\mathrm{s\,s}\!\bra{\uparrow} + {}_\mathrm{p}\!\!\braket{\phi|\theta}\!{}_\mathrm{p}\ket{\downarrow}_\mathrm{s\,s}\!\bra{\downarrow}$.

Based on these operators, we calculate the spin-down population after the photon count at the measurement basis $\ket{\phi}$, as shown in Fig.~5(a). These curves are obtained from the initial spin-down population of $P(\downarrow)=|\beta|^{2}=3/4, 1/2, 1/4$. When a photon is counted for the measurement basis $\ket{\phi}$ satisfying $\braket{\phi|0}=0$ ($\braket{\phi|\theta}=0$), the spin is projected to $\ket{\downarrow} (\ket{\uparrow})$. For the other measurement bases, the spin is weakly measured rather than projected. This implies that the spin-down population varies depending on the result of the measurement; however it is determined only probabilistically, which provides stochastic reversibility of the measurement~\cite{Royer94,Terashima06}. For example, successive measurements using opposite signs of rotation angles $\theta, -\theta$ and different measurement bases $\phi=\Delta, \Delta-\theta$ can stochastically restore the initial unknown spin state because of $\hat{M}(-\theta)_{\Delta-\theta}\hat{M}(\theta)_{\Delta}\propto\hat{I}$.
\begin{figure}[tbp]
\begin{center}
\scalebox{0.22}{\includegraphics{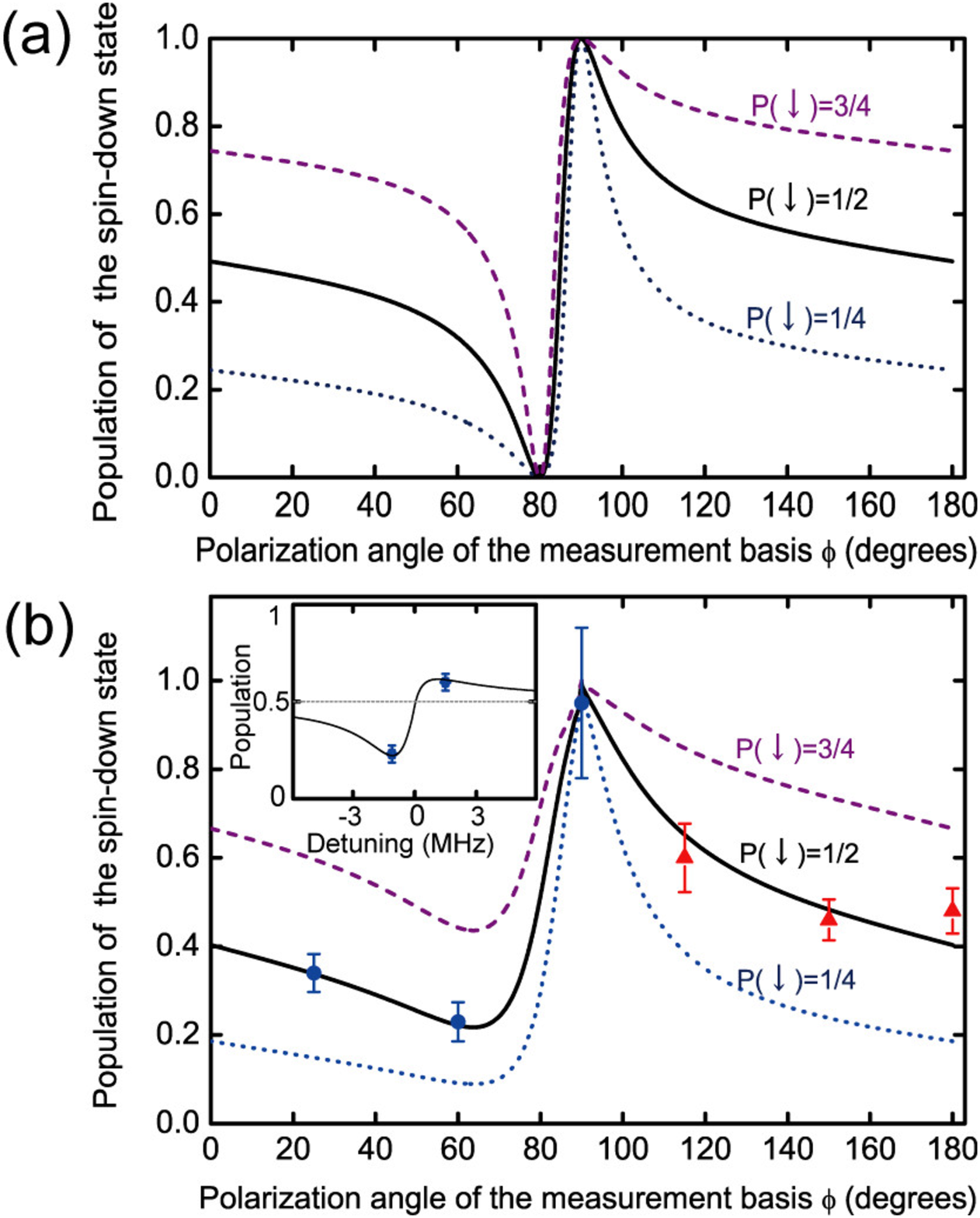}}
\caption{(color online)
(a) Population of the $\ket{\downarrow}$ state after the projection on the probe pulse as a function of the polarization angle $\phi$ of the measurement basis. The model includes only the pure rotation of polarization, and a rotation angle of $\theta=-10$~degrees for $\ket{\downarrow}$ is assumed. (b) Comparison of measured data with a theory that takes into account ellipticity. Blue circles show data taken conditioned on the detection of reflected photons. Red triangles show data  for transmitted photons. Note that if the rotation angle $\phi$ goes beyond $90$~degrees, the roles of the reflected and transmitted conditions interchange. The inset shows the population versus probe detuning for the basis of $\phi=60$~degrees.}
\label{action}
\end{center}
\end{figure}

In a real situation, the polarization of the transmitted probe exhibits ellipticity as well as rotation.
To treat the ellipticity, we calculate the conditional probabiliy, $P(\downarrow\!|\,\phi)$, of detecting the $|\!\!\downarrow\rangle$ state after the detection of a photon with polarization angle $\phi$.
According to Bayes's rule, the conditional probability is given by
\begin{equation}
P(\downarrow\!|\,\phi)=\frac{P(\phi|\!\downarrow)P(\downarrow)}{P(\phi|\!\uparrow)P(\uparrow)+P(\phi|\!\downarrow)P(\downarrow)}.
\end{equation}
Here, $P(\uparrow)$ and $P(\downarrow)$ are initial populations. $P(\phi|\!\downarrow)$ and $P(\phi|\!\uparrow)$ represent the probability of detecting a photon at angle $\phi$ when the atom is prepared in the $\ket{\downarrow}$ or $\ket{\uparrow}$ state, and they are given by
\begin{equation}
P(\phi|\!\downarrow) = \frac{|e^{-i\phi}T_-(\delta)+e^{+i\phi}T_+(\delta)|^2}{4}
\end{equation}
and $P(\phi|\!\uparrow) = \cos^2 \phi$.
By incorporating the effect of atomic motion into these probabilities, we obtain the theoretical curves in Fig.~5(b) which exhibit a tendency similar to those of the pure rotation cases shown in Fig.~5(a).

To demonstrate the tunability of the measurement process, the spin was first polarized along the x-direction, i.e., prepared in the superposition of $\ket{\uparrow}$ and $\ket{\downarrow}$ with $P(\uparrow)=P(\downarrow)=0.5$. For this purpose, after the coincidence, the spin was exposed to a spin-polarizing beam with the $^1\mathrm{S}_0$$-$$^1\mathrm{P}_1$ transition~(399~nm) from the x-axis in Fig.~1, as shown in Fig.~3~(b). 
Then, a 4 $\mu$s probe pulse with a detuning of -1.1~MHz and a mean photon number less than unity was incident onto the cavity, and the output was single photon-counted with the HWP rotated in a stepwise fashion.
Finally, the resultant variation in the population of the $\ket{\downarrow}$ state, which was initially 0.5, was measured by the spin projection method used in our previous work~\cite{Takeuchi09}.
For the projection, the excitation beam was switched on again. Due to the same reason as mentioned above in the initial spin-selection stage, the atom exposed to the excitation beam emits photons only when it is in the $|\!\!\downarrow\rangle$ state. If the number of counted photons is greater than zero during the measurement window of 20~$\mu$s, we consider the projection to the $|\!\!\downarrow\rangle$ state to be successfully carried out. Otherwise the spin is up, or the spin is down but the projection failed.

Points with error bars in Fig.~5(b) show the measured results of the conditional population of the $\ket{\downarrow}$ state, demonstrating that the population changes from the initial value of 0.5 after the photon counting of the probe pulse. The experimental data are in excellent agreement with numerically calculated curves. Note that the degree of the variation can be controlled by the angle of measurement basis $\phi$ and also by the probe detuning $\delta$ as shown in the inset of Fig.~5(b).

\section{Discussion}

We discuss here possible methods for achieving a larger Faraday rotation angle. In the context of realization of a controlled NOT gate, the amount of $45$~degrees rotation is sufficient for each spin state in our setup as mentioned above. Due to the bias magnetic field applied in the present work, however, the $\ket{\uparrow}$ state does not contribute to the Faraday rotation. By reducing the field sufficiently small, the $\ket{\uparrow}$ state rotates the polarization similarly. The amount of the rotation angle is the same as the $\ket{\downarrow}$ state, but the direction is opposite. We assume zero bias magnetic field hereafter.

Let us first consider the obstacles to the larger rotation angle in the present work. As shown in Fig.~4(b), the atomic motion inside the cavity was one of them, which made the rotation angle almost half of the calculated one with the maximum coupling rate $g_0$. If the tight confinement of the atoms inside the cavity is achieved~\cite{McKeever03,Nubmann05}, such motion would be greatly suppressed so that the rotation angle is expected to improve. Another source of the degradation was the condition of $\omega_p=\omega_c$ which was set for a technical reason. By stabilizing the cavity length with another beam at different wavelength from the resonance, which is usually done in modern optical cavity QED experiments~\cite{Mabuchi99}, further improvement is possible. By setting $\omega_c=\omega_a$ and scanning $\omega_p$ separately, the maximum rotation angle will rise to $23.6$~degrees at a specific detuning, while the maximum angle is $21.1$~degrees in the present work with $\omega_p=\omega_c$. Here we assume that the atoms stay at the anti-node of the cavity mode throughout the measurement. We also use this assumption in the following discussion.

\begin{figure}[tbp]
\begin{center}
\scalebox{0.22}{\includegraphics{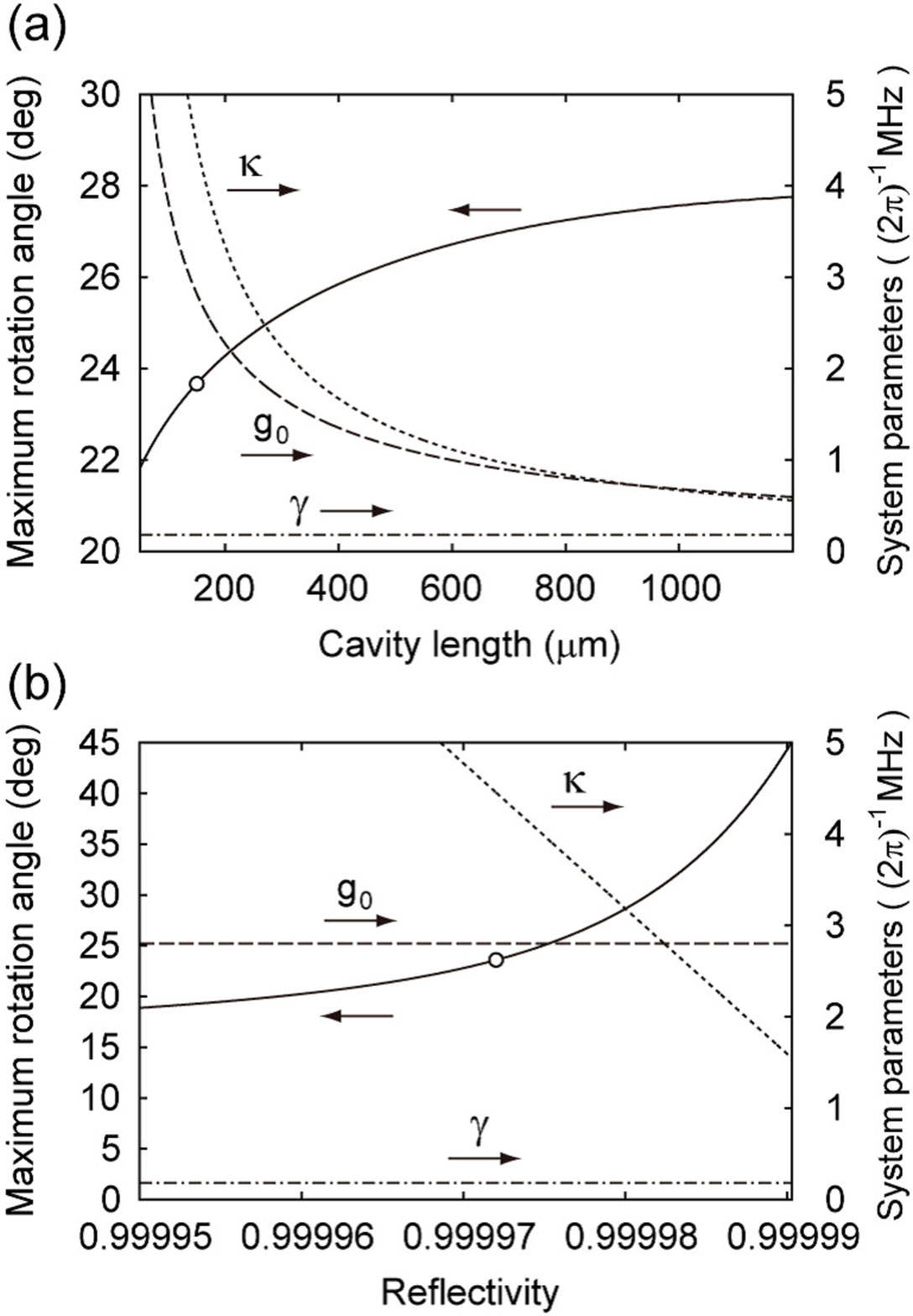}}
\caption{
Maximum rotation angle and the corresponding system parameters, $g_0$, $\kappa$, and $\gamma$, as a function of (a) the cavity length and (b) the reflectivity of the cavity mirrors. The open circles indicate the maximum angle of $23.6$~degrees that can be achieved in the present work with the cavity length 150 $\mu$m and the reflectivity 0.999972.
}\end{center}
\end{figure}
For the further improvement in the rotation angle, we next examine the following two methods: changing the cavity length and the reflectivity of mirrors that compose the cavity. These changes lead to the adjustment of relations among the system parameters describing our setup, namely $g_0$, $\kappa$, and $\gamma$, resulting in a variation of the available rotation angle. First we calculate the dependence of the maximum angle on the cavity length in Fig.~6(a). Because the rotation angle shows the dispersive behavior for the detuning $\omega_p-\omega_a$, we take the maximum value from the whole range of the detuning for each parameter space. By elongating the cavity, the maximum angle approached approximately $28$~degrees. Although the system reaches the strong coupling regime $g_0\ge\kappa, \gamma$, the improvement was modest. The dependence of the maximum angle on the reflectivity was estimated in Fig.~6(b). The higher the reflectivity is, the larger the rotation angle becomes. This is because the photons inside the cavity bounce more back and forth and interact more efficiently with the atoms. Accordingly the coupling rate $g_0$ becomes much larger than the cavity decay rate $\kappa$ as well as the atom decay rate $\gamma$, so that the system enters the strong coupling regime. At the reflectivity of 0.999990, the maximum angle was estimated to reach $45$~degrees, which is enough for implementing the controlled NOT gate.

\section{Conclusion}

We have observed Faraday rotation by an angle of more than $10$~degrees of photons caused by a single nuclear spin in a high-finesse cavity.
By projective measurements on the ancilla photon that transmitted through the cavity, we have realized projection measurement and weak measurement on the nuclear spin. 
Due to the short interaction time~$\simeq30~\mu$s, the measurements were limited to the populations, i.e. the diagonal terms in the density matrix.
However, by trapping an atom inside the cavity~\cite{McKeever03,Nubmann05}, it should be possible to probe off-diagonal terms by rotating the spin direction with an NMR method. Determination of both diagonal and off-diagonal terms will realize a full quantum state tomography of a single nuclear spin.

\bigskip

\begin{acknowledgments}
We acknowledge members of the ERATO macroscopic quantum control project for their useful discussions. Special thanks are due to T. Mukaiyama, T. Kishimoto, and S. Inouye for the fruitful discussions. We also appreciate R. Inoue for many stimulating discussions. One of the authors~(Y. E.) was partially supported by the JSPS.
\end{acknowledgments}


\begin{thebibliography}{99}

\bibitem{Julsgaard01} B. Julsgaard, A. Kozhekin, and E. S. Polzik,
Nature (London) \textbf{413,} 400 (2001).

\bibitem{Kuzmich98} A. Kuzmich, N. P. Bigelow, and L. Mandel,
Europhys. Lett. \textbf{42,} 481 (1998).

\bibitem{Takahashi99} Y. Takahashi \textit{et al.},
Phys. Rev. A \textbf{60,} 4974 (1999).

\bibitem{Takano09} T. Takano,M. Fuyama, R. Namiki, and Y. Takahashi,
Phys. Rev. Lett. \textbf{102,} 033601 (2009).

\bibitem{Kitagawa93} M. Kitagawa and M. Ueda,
Phys. Rev. A \textbf{47,} 5138 (1993).

\bibitem{Turchette95} Q. A. Turchette, C. J. Hood, W. Lange, H. Mabuchi, and H. J. Kimble,
Phys. Rev. Lett. \textbf{75,} 4710 (1995).

\bibitem{Aljunid09} S. A. Aljunid \textit{et al.},
Phys. Rev. Lett. \textbf{103,} 153601 (2009).

\bibitem{NC} M. A. Nielsen and I. L. Chuang, \textit{Quantum Computation and Quantum Information} (Cambridge University Press, Cambridge, 2000).

\bibitem{Hume07} D. B. Hume, T. Rosenband, and D. J. Wineland,
Phys. Rev. Lett. \textbf{99,} 120502 (2007).

\bibitem{Handel05} R. V. Handel, J. K. Stockton, and H. Mabuchi,
IEEE Trans. Automat. Control \textbf{50,} 768 (2005).

\bibitem{Ueda92} M. Ueda and M. Kitagawa,
Phys. Rev. Lett. \textbf{68,} 3424 (1992).

\bibitem{Royer94} A. Royer,
Phys. Rev. Lett. \textbf{73,} 913 (1994).

\bibitem{Terashima06} H. Terashima and M. Ueda,
Phys. Rev. A \textbf{74,} 012102 (2006).

\bibitem{Katz08} N. Katz \textit{et al.},
Phys. Rev. Lett. \textbf{101,} 200401 (2008).

\bibitem{Koashi99} M. Koashi and M. Ueda,
Phys. Rev. Lett. \textbf{82,} 2598 (1999).

\bibitem{Kane98} B. Kane,
Nature (London) \textbf{393,} 133 (1998).

\bibitem{Childress06} L. Childress \textit{et al.},
Science \textbf{314,} 281 (2006).

\bibitem{Takeuchi09} M. Takeuchi \textit{et al.},
arXiv:0907.0336 (2009).

\bibitem{Kleppner81} D. Kleppner,
Phys. Rev. Lett. \textbf{47,} 233 (1981).

\bibitem{Turchette95a} Q. A. Turchette, R. J. Thompson, and H. J. Kimble,
Appl. Phys. B \textbf{60,} S1 (1995).

\bibitem{Carmichael} H. Carmichael,
\textit{An open systems approach to quantum optics} (Springer-Verlag, 1993).

\bibitem{McKeever03} J. McKeever \textit{et al.},
Phys. Rev. Lett. \textbf{90,} 133602 (2003).

\bibitem{Nubmann05} S. Nu{\ss}mann \textit{et al.},
Nature Phys. \textbf{1,} 122 (2005).

\bibitem{Mabuchi99} H. Mabuchi, J. Ye, and H. J. Kimble,
Appl. Phys. B \textbf{68,} 1095 (1999).

\end{thebibliography}
\end{document}